\documentclass[12pt,aps,prd,preprint,showkeys]{revtex4}
\usepackage[utf8x]{inputenc}
\usepackage{amsmath}
\usepackage{amsfonts}
\usepackage{amssymb}
\usepackage{graphicx}

\begin{document}
\title{Magnetic reconnection in the wakes of cosmic strings}
\author{Dilip Kumar}
 \affiliation{School of Physics, University of Hyderabad, Gachibowli, Hyderabad, India 500046}
\author{Soma Sanyal}
\affiliation{School of Physics, University of Hyderabad, Gachibowli, Hyderabad, India 500046}

\begin{abstract}

The motion of cosmic strings in the universe leads to the generation of 
wakes behind them. We study magnetized wakes of cosmic strings moving in the post recombination plasma. We show that magnetic reconnection can occur in the post shock region. Since the width of the cosmic string wake is very small, the reconnection occurs over a very short lengthscale. The reconnection leads to a large amount of kinetic energy being released in the post shock region of the cosmic string wake. This enhances the kinetic energy released during the reconnection. We make a rudimentary estimate of the kinetic energy released by the magnetic reconnection in cosmic strings wakes and show that it can account for low energy Gamma Ray Bursts (GRB) in the post recombination era.


\end{abstract}

\maketitle

\section{Introduction}
The generation and evolution of magnetic fields have been studied quite extensively in the literature. There are many well established sources for the generation of these magnetic fields. One of the methods of generation of these fields involve the motion of cosmic strings. Cosmic strings are topological defects which are formed due to symmetry breaking phase transitions in the early universe \cite{vilenkin}. The motion of these strings through the cosmic plasma leads to a wake like structure behind them. Primordial magnetic fields in the early universe can be generated by these cosmic string wakes due to the Harrison mechanism as well as the Biermann mechanism. In most cases the seed magnetic field generated in the early universe then grows due to the dynamo mechanism.  

Various signatures of topological defects, especially cosmic strings have been predicted in the literature \cite{CMB2,stringsigna1}. Recently, the interaction of the shocks in the wakes of the cosmic string are being studied in detail to look for signatures of cosmic string wakes in the 21 cm redshift surveys \cite{stringsigna2}. There are many different kinds of strings based on the nature of the phase transition that lead to their formation. Some of the cosmological signals are specific to the kind of cosmic string being considered. For example, it has been shown that fast radio bursts (FRB) can be generated by cusps of superconducting strings (SCS)\cite{vachaspati,zadorozhna,yu}. Apart from this, cusps of superconducting strings can also serve as engines for generating Gamma Ray Bursts (GRB's) \cite{berezinsky}. These are all shown to be signatures of a superconducting string network in the early universe. 


In this work, we show that Gamma Ray Bursts can result from magnetic reconnection in the wake of an Abelian Higgs strings. The GRB's occur when the magnetic field lines come close together in the cosmic string wake and release a large amount of energy through magnetic reconnection. In this case, the cosmic string need not be a superconducting cosmic string. Magnetic fields can be generated in the wake of Abelian Higgs string during the radiation epoch by the Biermann battery mechanism \cite{sovan}. The Biermann mechanism generates the field when there are density inhomogeneities in the wake due to the motion of neutrinos around the cosmic string. The neutrinos and the electrons interact through the weak ponderomotive force and an inhomogeneity is generated in the electron distribution in the wake of the cosmic string. It is well known that there is a temperature  difference across the wake due to the shock structure \cite{layek}. The electron inhomogeneity generated by the neutrinos however is not aligned to the temperature gradient. This misalignment of the electron density fluctuation and the temperature gradient results in the generation of a magnetic field.   
If these fields are enhanced due to the turbulence in the plasma, they will survive till the matter epoch. Our previous (numerical) work \cite{soumen} has indicated that there is a possibility of magnetic reconnection in cosmic string wakes, in this work we show how magnetic reconnection  occurs in the wake of cosmic strings and discuss the possibility that they may lead to Gamma Ray Bursts in the post recombination era. 

Magnetic reconnection occurs in highly conducting plasmas. In this process, the field lines of oppositely directed magnetic fields come close together and rearrange their topology by breaking and reconnecting.  
In this process, energy in the magnetic field is released to the surrounding plasma. Depending upon the magnitude of the magnetic field and the length of the reconnection region, the energy released can be very large. The process of magnetic reconnection is a complex process which involves the motion of charged particles and the fields generated by these charged particles over a short lengthscale and timescale. The first model to describe the process of reconnection in detail was the Sweet-Parker model \cite{sweetparker}. The model gave an estimate of the energy released during the magnetic reconnection process. 
Later several other models were developed to understand the process in more detail \cite{SPreconnection}. In this work, we have used the Sweet Parker model to make an initial estimate of the energy that can be released by magnetic reconnection in cosmic string wakes. Though the energy released depends on the magnetic field, we find that a conservative estimate shows that the energy released may lead to a Gamma Ray Bursts(GRB).     

GRB's are large explosions of energy over a finite period of time. These may be short duration bursts of very high energy or longer duration bursts of lower energies. The energy emission mechanism for a GRB is still not well understood. Most of the GRB's are attributed to the merger of two neutron stars or to the collapse of a star. Some of the GRB's are also attributed to the process of magnetic reconnection in the plasma. GRB's from magnetic reconnection have previously been studied in the context of a turbulent magnetic plasma with a high magnetization parameter \cite{granot}. Detailed numerical simulation are currently being done to investigate the possibility of generating GRB's due to magnetic reconnection around jets and collapsing neutron stars. In this work, we show that GRB's can also be emitted from magnetic reconnection in cosmic string wakes.

In section II we revise the formation of shocks in cosmic string wakes and the generation of magnetic fields in the shock.  In section III, we discuss the possibility of magnetic reconnection occurring in these wakes. In section IV, we present the cosmological consequences of the magnetic reconnection occurring in these wakes and predict the possibility of observational signals arising from these reconnections. Finally we present our conclusions in section V. 

\section{Shocks and magnetic fields in cosmic string wakes}

As mentioned in the introduction, cosmic strings generated from symmetry breaking phase transitions in the early universe, produce a wake like structure behind them. The wake like structure is due to the conical spacetime of the cosmic string. Shocks are generated in the wakes of these strings as they move through the plasma. A detailed analysis of the shock structure can be obtained from the following references \cite{shocks}. In a recent work, it was shown that a magnetic field can be generated in the shock of a cosmic string due to the Biermann mechanism \cite{sovan}. We briefly review the generation of the magnetic field in the shock in this section and show in the next section why such magnetic fields will lead to magnetic reconnections in the wake of a cosmic string.

Massive particles moving around cosmic strings often form closed orbits around the string \cite{abhisek}. If these  particles happen to be neutrinos then they generate a neutral current close to the string. The neutrinos interact with the electrons in the plasma by the weak force and cause electron currents in the plasma. These currents lead to oscillatory density fluctuations of the electron density in the wake of the cosmic string. It was shown in ref \cite{sovan} that these density fluctuations can lead to the generation of a magnetic field in the cosmic string wake through the Biermann mechanism.  

The evolution of the magnetic field by the Biermann battery mechanism is given by, 
\begin{multline}
\frac{\partial \vec{B}}{\partial t} = \nabla \times (\vec{v_e} \times \vec{B}) + \frac{\eta_{res}}{4 \pi} \nabla^2 \vec{B} - \frac{1}{e N_e } 
 \nabla \times (\vec{j} \times \vec{B}) \\ 
 - \frac{1}{N_e e} \nabla N_e \times \nabla T
 \end{multline} 
Here, $\vec{B}$ is the magnetic field generated by the electron current, $\vec{v_e}$ is the velocity of the electrons in the plasma,$\eta_{res}$ is the resistivity of the plasma, $e$ is the electric charge, $N_e$ is the electron number density and $T$ is the temperature of the plasma. In the absence of a magnetic field in the plasma, it is the last term which generates the magnetic field. 

We will now establish that due to the temperature difference across the shock wave, the Biermann term (last term) will generate opposite magnetic fields on the two sides of the shock. The density gradient in the cosmic string wake is non uniform and the gradient of the electron number density is given by, 
\begin{equation}
\nabla N_e = \hat{i} \frac{\partial N_e}{\partial x} + \hat{j} \frac{\partial N_e}{\partial y} + \hat{k}\frac{\partial N_e}{\partial z}
\end{equation}

In the case of the cosmic string wakes, the temperature gradient is perpendicular to the flow direction. Though we can have small fluctuations in other direction the gradient of the temperature is dominated by $\hat{j} \frac{\partial T}{\partial y}$. If we consider the origin of the coordinate system to be at the position of the cosmic string then the two sides of the planar shock generated by the moving cosmic string will have opposite temperature gradients. Figure 1 illustrates the concept. The upper half of the shock is referred to as "A" while the lower part is referred to as "R". Therefore, if the temperature gradient at A is $\hat{j} \frac{\partial T}{\partial y}$
then at R it will be  $-\hat{j} \frac{\partial T}{\partial y}$. 
The number density gradient remains the same at both A and R sides. This means that the magnetic field generated by the Biermann mechanism at A and R are equal and opposite. 
\begin{figure}
\includegraphics[width = 86mm]{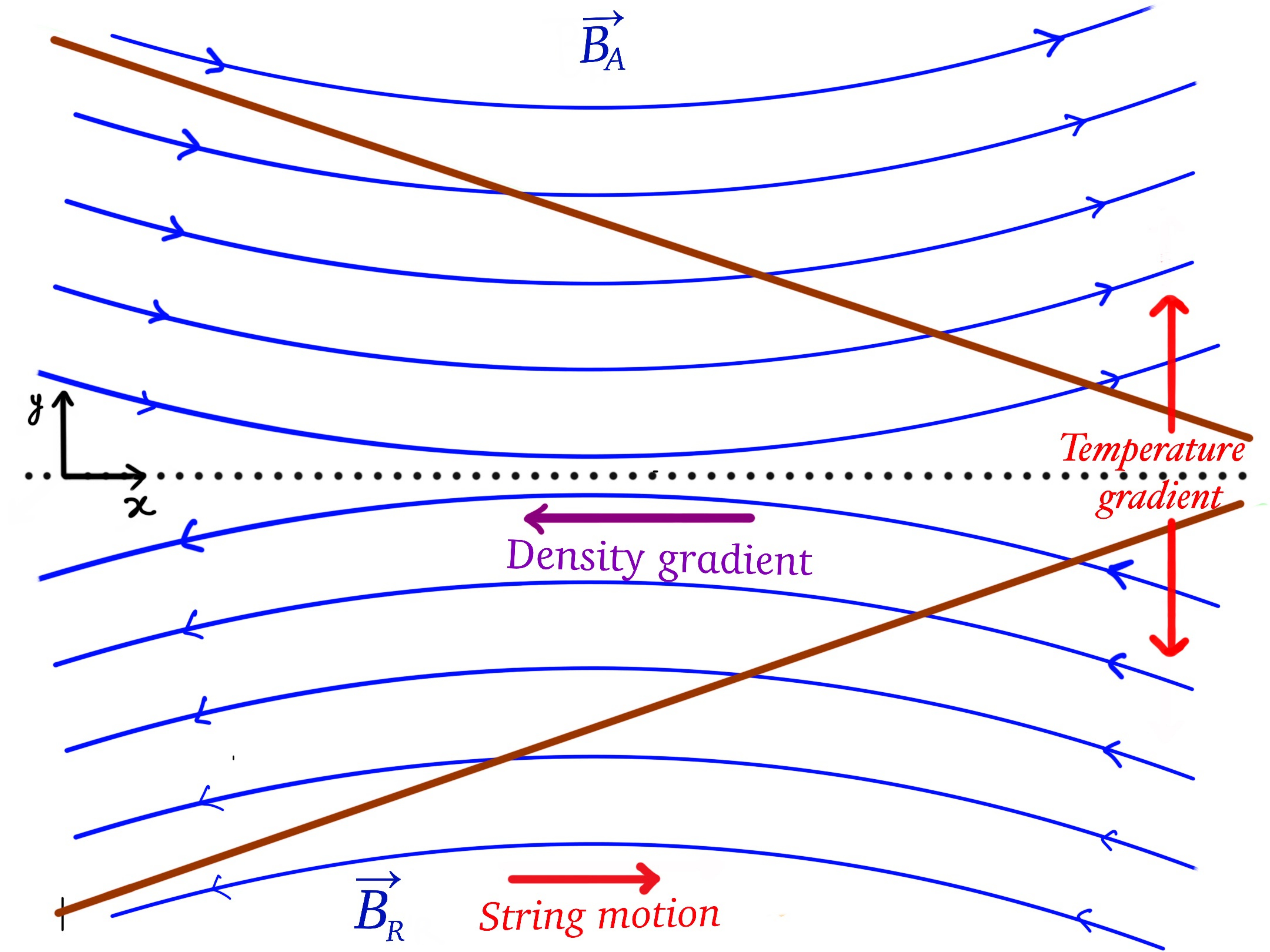}
\caption{Schematic diagram of the magnetic field lines generated by the cosmic string as it moves through the plasma. The brown (solid)  line gives the outline of the cosmic string wake and the blue (grey) lines are the directions of the fields that are generated due to the Biermann mechanism.}
\end{figure}
\begin{equation}
\vec{B}_A ~=~ -\hat{i} \frac{\partial N_e}{\partial z} \frac{\partial T}{\partial y} ~+ ~\hat{k} \frac{\partial N_e}{\partial x} \frac{\partial T}{\partial y}~ = ~-\vec{B}_R
\end{equation}
     
 Since the plasma is highly conducting, the magnetic Reynolds number is very large. This makes the magnetic diffusivity very large. The two magnetic field lines then diffuse and squeeze the fluid like two infinitely conducting sheets. We thus have a situation that was first described by Sweet and Parker \cite{SPreconnection} in their model for magnetic reconnections in conducting plasmas. This means that there is a strong possibility that the magnetic reconnections will occur in the shocks generated in the wakes of cosmic strings.  

In recent simulations of cosmic string wakes, lines of magnetic reconnection have been observed as the string moves through the plasma. These magnetic reconnection lines have been observed for a magnetic field in the wake of a cosmic string with some spatial variations. In the next section, we calculate the reconnection rates and length of the charge sheets that can be generated in the wakes of cosmic strings based on the Biermann mechanism outlined previously in ref \cite{sovan}.

\section{Magnetic reconnection in the wakes}

According to the Sweet-Parker model of magnetic reconnection \cite{sweetparker}, a current layer is formed between two oppositely directed magnetic fields generated in a conducting plasma. The electrical current density in these regions becomes very high. This forces the magnetic lines of force to break and reconnect, changing the topology of the magnetic field. The curvature force associated with the changed configuration results in the conversion of magnetic energy to kinetic energy. 

Assuming steady state conditions, the length of the diffusion region  between the two oppositely directed diffusing fields is given by, 
$l = \frac{u L }{v}$
here $u$ is the velocity with which the field lines merge and $v$ is the velocity at  which the fluid is expelled from the diffusion region. $L$ is the scale of the magnetic field that we are considering. In the cosmic string wake, $L$ will be determined by the dimensions of the wake. 
We are looking at the motion in two dimensions. Let us assume that the string is moving along the $x$ direction. The lines of forces will be generating a strain in the wake of the string. We assume that the magnetic field changes very slowly in the $y$ direction, so that we can assume the $y$ component of the magnetic field to be approximately constant. We are then left with the $x$-component of the magnetic field. The field lines push in the $y$ direction and the pressure difference due to the streching of the fields leads to the velocity of the diffusing fields.  
The velocity of the diffusing  field lines  will be given by, 
$u = \frac{c^2}{ 4 \pi \sigma B_x} (\frac{\partial B_x}{\partial y})$, 
here $\sigma $ is the conductivity of the fluid. The quantity $(\frac{\partial B_x}{\partial y})$ is evaluated at $y = 0$, this is the Sweet Parker model. 
The fluid is expelled with the characteristic velocity $v$ given by, 
$v \sim \frac{B}{(4 \pi \rho)^{1/2}}$,
this is known as the hydromagnetic velocity and it will be henceforth denoted by $C_H$, $\rho$ is the mass density of the plasma.  As can be seen from the expression, this is equivalent to the Alfven velocity in the magnetized wake.  In fig 1,  we have briefly sketched the magnetic field generated by the wakes and shown how it can be mapped onto the Sweet Parker model.  

The length of the diffusion region depends on the dimensions of the cosmic string wake. For a long single string generated at $t_f \le t_{eq}$, the dimensions of the wakes in terms of the redshift are given by $l_f$ and $w_f$. 
Here $l_f = t_f \frac{z_f}{z}$ and $w_f = v_s \gamma_s  t_f \frac{z_f}{z}$. Here, the suffix $f$ stands for the time at which the string is generated and the suffix $eq$ stands for the time after which the different particles went out of equilibrium.  The thickness of the wake is given by $ 4 \pi G \mu \gamma_s v_s t_i  \frac{z_f}{z} $, where $t_i$ is the time at which the reconnection occurs.  According to the Sweet Parker model of magnetic reconnection, the magnetic energy is converted into the kinetic energy of the plasma. The sudden increase in the kinetic energy leads to the acceleration of the particles in the plasma. Thus an estimate of the velocity of the merging fields gives us an estimate of the increase in the 
kinetic energy of the plasma. For this we calculate  $(\frac{\partial B_x}{\partial y}) $.    The details of the derivation can be obtained from \cite{sweetparker}. The final result is, 
$K.E = \frac{1}{2} \rho c^2 \left( \frac{C_H}{\sigma L} \right)$
where $c$ is the velocity of light. So, a large amount of kinetic energy is released for a dense plasma with a low electrical conductivity and a high magnetic field. 
For a cosmic string wake, the density is usually four times the density of the background plasma. Though the magnetic field in the wake may not be very high it may be high enough to generate a  considerable amount of energy which could have prospects of detection by various methods.

As the cosmic string moves through the plasma, the emission of kinetic energy can result in a burst of electromagnetic radiation. Since the wake behind the cosmic string is quite narrow, this can be detected as a burst of gamma rays. In the next section, we calculate the kinetic energy that can be released due to the reconnection in the wakes of the cosmic strings.

\section{Gamma Ray Bursts from magnetic reconnection}

The release of a large amount of energy in the plasma accelerates the particles in the plasma which may lead to the generation of a significant amount of radiation. So, it is quite possible that magnetic reconnection in the shocks of cosmic strings can lead to a burst of energy that could be a Gamma Ray Progenitor. There exist some GRB models which show magnetic reconnection to be the progenitor of a GRB but this is the first time that magnetic reconnections in the cosmic string wakes has been considered as the progenitor of a GRB. A detailed numerical calculation is beyond the scope of this work so we do an order of magnitude estimate of the energy that is released due to magnetic reconnection in the wake.

The kinetic energy that is released due to reconnection depends inversely
 on the lengthscale of the reconnecting magnetic field. Since the opening
  angle of the cosmic string wake is very small, this lengthscale is also
 very small. It will depend on the thickness of the wake and would be of
  the order of $ 4 \pi G \mu \gamma_s v_s t_i  \frac{z_f}{z} $. One of 
  the reasons why magnetic reconnection are traditionally not considered for GRB's is because of the large lengthscale over which they occur. In typical astrophysical plasma, the lengthscale over which reconnection occurs is considerably large, however in the case of cosmic string wakes, only the length of the wake is large, the thickness of the wake is much smaller as it is determined by the deficit angle of the cosmic string. This depends on the symmetry breaking scale. Since the GUT symmetry breaking phase transition happened at $10^{16}$GeV, the deficit angle of the GUT cosmic string is typically of the order of $10^{-12}- 10^{-6}$. Abelian Higgs strings have $G \mu$ of the order of $10^{-11}- 10^{-8}$. Current observational limits from the Planck collaboration have put the upper limit of Abelian Higgs strings at $3 \times 10^{-7}$ \cite{planck}. The order of the string velocity is $< \gamma_s v_s > \sim 0.4$. The time of the wake formation can be written as, $t_f = \left(\frac{3}{2} H(z) \sqrt{\frac{z_f+1}{z+1}} \right) $. Assuming that the wake formation happened around the recombination epoch $(z \sim 3000)$ and the magnetic reconnection is happening at lower values of the redshift $(z \sim 30)$, the lengthscale turns out to be $10^{-8}$ in dimensionless units. The conductivity of the plasma in the matter dominated epoch is given by $\sigma \sim 10^{10} s^{-1}$ \cite{caprini}. The kinetic energy is thus dependent primarily on the magnetic field in the wake and the plasma density. The plasma density at recombination is given by $3.8 \times 10^{-3} eV^{4}$. Since the value of $C_H$ depends upon the magnetic field and is generally greater than one for strongly magnetized plasma, as a first approximation we consider it to be of the order of one. An approximate estimate will then put the kinetic energy generated due to the magnetic reconnection to be of the order of $10^{14} erg/cm^2$. 
   
Since there have been GRB's detected in the lower energy range $(10-10000)$ keV which have fluences of the order of $10^{-5} erg/cm^2$, it is quite possible that the magnetic reconnection in magnetized wakes of cosmic strings can be detected as a GRB. Our rudimentary method unfortunately does not allow us to determine the timescale of the GRB which would have been helpful in identifying an observational signal in a better way. However, a large number of GRB's have been reported with flux densities ranging from $10^{-5} erg/cm^2$ to $2 X 10^{-4} erg/cm^2$, with burst durations ranging from less than $0.1$ sec  to $30$ secs \cite{klebesadle}. As the time duration is also similar to the timescales observed in magnetic reconnections, we plan to look into these GRB's in greater detail to see if they can be signatures of magnetized cosmic string wakes.

\section{Conclusions}
In this paper, we have shown that it is possible to generate GRB's from magnetic reconnection in the wakes of Abelian Higgs cosmic strings. Magnetic reconnections are currently being studied in both laboratory and astrophysical plasmas. It is manifested in solar flares, coronal mass ejection and other solar phenomenon. The search for signatures of cosmic strings and their wakes are also continuing through both direct detection methods as well as statistical methods. Cosmic string networks are known to generate both gravitational and non gravitational signatures. The signatures related to magnetic fields and electromagnetic phenomenon have however been limited to superconducting cosmic strings. Our work shows that in any cosmic string wake if the magnetic field evolves by the Biermann mechanism, it is possible to have magnetic reconnections. All it requires is a density inhomogeneity which is not aligned to the temperature gradient of the cosmic string wake. Though in this work we have used a specific method to show the phenomenon analytically, it may be extended to many other cases where magnetic fields occur in cosmic string wakes.  We have only done a preliminary calculation for the energy of the GRB. An important parameter which distinguishes a GRB is the duration of the burst. This helps in the classification of the observed GRBs. In this current work we have not estimated the duration of the emitted GRB in our model.

Since this is a preliminary work, we have not been able to obtain an exact match to the observed GRB's. We have used the Sweet-Parker model, and found that the energy generated by the reconnection is quite significant. There are various improvements on the Sweet-Parker model.  We are currently working on a more detailed model that will tell us more about the reconnection phenomenon in the magnetized wakes of cosmic strings.

\begin{center}
 Acknowledgments
\end{center}
The authors D.K and S.S would like to thank Biswanath Layek for the interesting discussions which helped to improve the paper considerably.This work is supported by the DST- SERB Power Grant no. SPG/2021/002228 of the Govt. of India.

\end{document}